\documentclass[aps,pra,groupedaddress]{revtex4}
\usepackage[british]{babel} %
\usepackage[latin1]{inputenc} 
\usepackage[a4paper,top=3cm ,left=3.5cm,right=2.5cm,bottom=2.5cm]{geometry}
\usepackage{amssymb}
\usepackage{amsmath, amsthm}
\usepackage{epsfig}
\usepackage{hyperref}
\usepackage[none]{hyphenat}
\usepackage{tikz}
\usepackage{verbatim}

\newtheorem{teo}{Theorem}[section]
\newtheorem{cor}[subsection]{Corollary}

\newtheorem{prop}[teo]{Proposition}
\newtheorem{defi}[teo]{Definition}

\newcommand{\R}{\mathbb R}

\newcommand{\norm}[1]{\left\Vert#1\right\Vert}
\newcommand{\abs}[1]{\left\vert#1\right\vert}
\newcommand{\set}[1]{\left\{#1\right\}}

\newcommand{\eps}{\varepsilon}

\newcommand{\Rn}{\mathbb{R}^n}

\newcommand{\trace}[1]{\mbox{tr}\left( #1 \right)}
\newcommand{\goes}{\rightarrow}

\newcommand{\ket}[1]{\vert#1\rangle}
\newcommand{\bra}[1]{\langle#1\vert}
\newcommand{\braket}[2]{\langle#1\vert#2\rangle}

\newcommand{\inter}{\cap}

\newcommand{\refe}[1]{(\ref{#1})}

\newcommand{\interior}[1]{\mbox{int}\left({#1}\right)}

\usepackage{graphicx}
\usepackage{amsmath}
\usepackage{amssymb}
\usepackage{bm}

\graphicspath{{converted_graphics/}}
\begin{document}
\title{Local solutions of Maximum Likelihood Estimation \\ in Quantum State Tomography}


\author{Douglas S. Gon\c{c}alves}
\affiliation{Departamento de Matem\'atica Aplicada, Universidade Estadual de Campinas,
Campinas, SP 13083-859, Brazil}
\author{M\'arcia A. Gomes-Ruggiero}
\affiliation{Departamento de Matem\'atica Aplicada, Universidade Estadual de Campinas,
Campinas, SP 13083-859, Brazil}
\author{Carlile Lavor}
\affiliation{Departamento de Matem\'atica Aplicada, Universidade Estadual de Campinas,
Campinas, SP 13083-859, Brazil}
\author{Osvaldo Jim\'enez Far\'ias}
\affiliation{Instituto de Ciencias Nucleares, Universidad Nacional Aut\'onoma de M\'exico
(UNAM), Apdo. Postal 70-543, M\'exico 04510 D.F.}
\affiliation{Instituto de F\'{\i}sica, Universidade Federal do Rio de
Janeiro, Caixa Postal 68528, Rio de Janeiro, RJ 21941-972, Brazil}
\author{P. H. Souto Ribeiro}
\email{phsr@if.ufrj.br}
\affiliation{Instituto de F\'{\i}sica, Universidade Federal do Rio de
Janeiro, Caixa Postal 68528, Rio de Janeiro, RJ 21941-972, Brazil}
\begin{abstract}
Maximum likelihood estimation is one of the most used methods in quantum state tomography, where the aim is to reconstruct the density matrix of a physical system from measurement results.  One strategy to deal with positivity and unit trace constraints is to parameterize the matrix to be reconstructed in order to ensure that it is physical. In this case, the negative log-likelihood function in terms of the parameters, may have several local minima. In various papers in the field, a source of errors in this process has been associated to the possibility that most of these local minima are not global, so that optimization methods could be trapped in the wrong minimum, leading to a wrong density matrix. Here we show that, for convex negative log-likelihood functions, all local minima of the unconstrained parameterized problem are global, thus any minimizer leads to the maximum likelihood estimation for the density matrix. We also discuss some practical sources of errors. 
\end{abstract}
\maketitle
\section{Introduction}\label{intro}
Preparation, manipulation and characterization of quantum systems are essential tasks for quantum information processing \cite{james2001, nc2000, clavor2011}. It is known that measurement on a quantum system projects it onto a probable state. Therefore it is impossible to determine the state of the system with a single copy. Moreover the non-cloning theorem \cite{wz1982} forbids the production of several copies of the unknown state for its further reconstruction. However, if one has a source of quantum systems which are identically prepared, it is possible to reconstruct its quantum state through Quantum State Tomography (QST): an experimental process where the ensemble of unknown, but identically prepared quantum states, is characterized by a sequence of measurements
in different basis, allowing the reconstruction of its density matrix.

The first approach that appeared in literature was linear inversion \cite{vr1989}, based on inversion of the
Born's rule, where the probabilities predicted by quantum mechanics are equated to experimental normalized
frequencies. However, due to experimental noise, the recovered state may not correspond
to a physical state. To ensure that the estimated density matrix is physical, we need to restrict 
it to the space of semidefinite positive and unit trace matrices. This is a problem of statistical inference called constrained parameter estimation \cite{cb2001,lc1998}.

The most commonly used method for parameter estimation, due to its good asymptotic properties \cite{lc1998,cb2001}, is the maximum likelihood estimation (MLE).
It was used in QST \cite{hradil1997}, and in particular for the reconstruction of states of two photonic qubits \cite{james2001}. This method has 
been widely used and discussed in the literature \cite{kaznady2009, rehacek2007, blume2010}. Other methods can also be used for QST, for instance the ones based on Bayesian Inference, as discussed in \cite{blume2010}. However, in this work we focus on maximum likelihood estimation.

The MLE method seeks the matrix that maximizes the probability of observed experimental data, restricting the search to semidefinite positive and unit trace density matrices. Therefore, to get the maximum likelihood estimation, we need to solve a constrained optimization problem \cite{luenberger2003, nocedal1999}. As the dimension of the matrix to be reconstructed increases, the optimization problem becomes more difficult. Alternative approaches to full optimization have been discussed \cite{kaznady2009}, but assuming some prior knowledge about the state. Other methods to obtain the MLE of density matrix are based on extremal equations, like $R \rho R$ iteration presented in \cite{rehacek2007}. In this work, we study the parameterization approach that appeared in \cite{james2001,altepeter2005} and show the equivalence between local solutions of the associated unconstrained optimization problem.

One way of treating the problem of the constraints is parameterizing the matrix to ensure positivity and unit trace, and solving the new unconstrained optimization problem with these parameters\cite{james2001}. However, this new optimization problem has a nonconvex objective function with several local minima. Since MLE method requires the global minimum, we could have a problem, because a local non-global minimum could lead to a wrong density matrix estimation. These local minima were taken as possible non-global in the literature \cite{james2001,altepeter2005,usami2003} and this problem was pointed as a source of errors in quantum state reconstruction. Our contribution here is to show that all local minimizers are in fact global. Therefore, no matter what minimum the optimization method finds, the reconstructed density matrix is always the same. We also show that some complaints found in the literature, about errors in the reconstruction, may be caused by the use of optimization methods where global convergence \cite{luenberger2003, nocedal1999} to local minimizers does not hold or stop prematurely, indicating the stagnation (poor progress) of the optimization process.

This paper is organized as follow. Section \ref{limle} reviews the standard methods of Quantum State Tomography: linear inversion and maximum likelihood estimation (MLE). Section \ref{parameterization} discusses the parameterization of the density matrix in MLE to assure that the reconstructed matrix is physical. The mathematical foundation needed to prove the equivalence of local minimizers for the unconstrained optimization problem is presented in Section \ref{theory}. Section \ref{main} presents our main result as a corollary of the theorem demonstrated in Section \ref{theory}. The spurious local minima, for the associated unconstrained optimization problem, arise from over-parameterization of the density matrix, but we show that all of them return the same global minima and the same reconstructed density matrix. Some examples illustrating this point and some problems that may occur using inappropriate optimization procedures are presented in Section \ref{num_ex}. Finally, we conclude in Section \ref{conclusion}.

\section{Linear Inversion and MLE}\label{limle}
In this section we review the basic ideas of linear inversion and maximum likelihood estimation.

\subsection{Linear Inversion}

Let $\set{\hat{\Gamma}_{\nu}}$ be an orthonormal basis for Hermitian matrices space of order $d$, where
\begin{eqnarray}
\trace{ \hat{\Gamma}_{\nu} \hat{\Gamma}_{\mu}} = \delta_{\nu,\mu}\ \ ,
\\ \nonumber
\\ \nonumber
A = \sum_{\nu=1}^{d^2} \trace{\hat{\Gamma}_{\nu} A} \hat{\Gamma}_{\nu}\ \ ,\ \  \forall A.
\end{eqnarray}
Using this basis, we can write a density matrix $\rho$ as
\begin{equation}
\rho =  \sum_{\nu=1}^{d^2} S_{\nu}\Gamma_{\nu},
\label{rhogamma}
\end{equation}
\noindent where the coefficients $S_{\nu} = \trace{\Gamma_{\nu}\rho}$ are known as Stokes parameters.
\noindent If we take measurements described by a POVM set $\set{O_{\mu}}$, quantum mechanics tells us that
\begin{equation}
f_{\mu} \approx p(\mu) = \trace{O_{\mu}\rho},
\label{expectation}
\end{equation}
\noindent where $f_{\mu} = n_{\mu}/N$ denotes the normalized frequencies, ($n_{\mu}$ and $N$ denote the number of times outcome $\mu$ occurs and the normalization constant respectively.). Taking the inner product using  \refe{rhogamma} and \refe{expectation}, we have
\begin{equation}
f_{\mu} \approx \trace{O_{\mu}\rho} =  \sum_{\nu} S_{\nu}\  \trace{O_{\mu}\Gamma_{\nu}}.
\label{equacoes}
\end{equation}
\noindent Consider the index $\mu$ labelling the experimental setup/result pairs. Then taking an informationally  complete set of measurements, we can determine the parameters $S_{\nu}$, and therefore $\rho$, solving the linear system
\begin{equation}
Bs = f,
\label{sl}
\end{equation}
\noindent where $s$ is the vector with $S_{\nu}$ components, $f$ is the vector with components $f_{\mu}$, and $B$ is the $d^2 \times d^2$ matrix with entries $B_{\nu\mu} = \trace{O_{\mu}\Gamma_{\nu}}$. This method is called linear tomography reconstruction or linear inversion. \\
Nevertheless, the experimental noise disturbs the normalized frequencies $f_{\mu}$, which in turn do not produce good approximations for probabilities $\trace{O_{\mu}\rho}$. Moreover, the law of large numbers states that the normalized frequencies approximate the theoretical probabilities only if the measurements are repeated a large number of times, which is not always the case. This implies that the recovered matrix may not be a legitimate density matrix, having negative eigenvalues or not having unit trace. In order to overcome this problem, it is usual to apply the maximum likelihood estimation method \cite{james2001,hradil1997}. 

\subsection{Maximum Likelihood Estimation}

In statistical inference \cite{cb2001,lc1998}, the Maximum Likelihood Estimation (MLE) is a method to estimate an unknown parameter $\theta$ of a population, based on sampled data~$x$. The idea is to maximize the conditional probability $P(x\,|\, \theta)$, that is, the maximum likelihood estimation $\hat{\theta}_{mle}$ is the value that maximizes the probability of getting the observed data.\\
The MLE has a lot of good properties \cite{lc1998} from the statistical point of view. This explains why it is widely used for point estimation. 
Among others, we can cite: \\
\begin{itemize}
\item {\bf Consistency}: $\displaystyle \lim_{n \rightarrow \infty} P(|\hat{\theta}_n - \theta|<\eps)=1, \forall \eps>0$, where $\hat{\theta}_n$ is the estimation for a finite sample of size $n$. This means that the maximum likelihood estimator $\hat{\theta}$ is unbiased when $n \goes \infty$.

\item {\bf Asymptotic normality and efficiency}: as $n \goes \infty$, $\ \hat{\theta} \sim N(\theta,{\cal I}(\theta)^{-1})$, that is, $\hat{\theta}$ has a normal distribution with mean $\theta$ and covariance matrix ${\cal I}(\theta)^{-1}$, where ${\cal I}(\theta)$ is the Fisher information matrix. In other words, as the sample size increases the MLE becomes an efficient estimator.

\item In a {\bf finite sample}, if a minimum variance unbiased estimator exists, then the method of MLE chooses it; that is, MLE performs at least as well as competitors in finite samples.

\item Finally, the {\bf invariance} property, that is, if $\hat{\theta}$ is the MLE for $\theta$ then $\tau(\hat{\theta})$ is the MLE for $\tau(\theta)$. Note that, for any observable $O$, if $\hat{\rho}$ is the MLE for $\rho$, then $\trace{O\hat{\rho}}$ is the MLE for $\trace{O\rho}$.

\end{itemize}
\noindent In QST \cite{altepeter2005,hradil1997,james2001}, the MLE can be formulated as

\begin{equation}
\begin{aligned}
\max_{\rho} \ \ \ & {\cal L}(\rho) \\
\text{s.t} \ \ \ & \trace{\rho} \ = 1, & \  \\
& \rho \ \succeq 0, & 
\end{aligned}
\label{mle}
\end{equation}

\noindent where ${\cal L}(\rho) = P(n | \rho)$ is the probability of getting the experimental outcomes $n$ given the parameter $\rho$. The function ${\cal L}(\rho)$ is called likelihood function.\\
\\
Usually, instead of maximizing ${\cal L}(\rho)$, it is easier to minimize the negative log-likelihood function 
$$
F(\rho) = - \log {\cal L}(\rho).
$$
So, the optimization problem is
\begin{equation}
\begin{aligned}
\min_{\rho} \ \ \ & F(\rho) \\
\text{s.t} \ \ \ & \trace{\rho} \ = 1, & \  \\
& \rho \ \succeq 0. & 
\end{aligned}
\label{logmle}
\end{equation}
From the optimization point of view, if $F(\rho)$ is a convex function, we have a convex optimization problem \cite{boyd2004} since the feasible set ${\cal S} = \set{\rho \ | \ \rho = \rho^ {\dagger},\ \trace{\rho}=1,\ \rho \succeq 0 }$ is convex. This kind of problem is interesting because every local minimum is a global one.\\
The constraints defining the optimization problem \refe{logmle} are known as linear matrix inequalities/equalities \cite{boyd2004} and when the objective function $F(\rho)$ is linear we have a semidefinite programming problem (SDP) \cite{boyd2004,todd2001} but, in general, this is not the case.\\
\\
Although we have efficient interior point methods for semidefinite programming \cite{boyd2004,todd2001}, the reformulation of \refe{logmle} into a linear SDP form requires the introduction of auxiliary variables and constraints that increase the dimension, and therefore the difficulty of the optimization problem. \\
\\
Here we consider a reformulation presented in \cite{james2001,altepeter2005} that converts the problem \refe{logmle} into an unconstrained optimization problem. 

\section{Parameterized Density Matrix}\label{parameterization}
Let us remember that the matrix $\rho$ must be Hermitian semidefinite positive and have unit trace. To satisfy the non negativity constraint, we can consider the product $T^{\dagger}T$, for any matrix $T$, where it is clear that
\begin{equation}
\label{T}
\bra{\psi} T^{\dagger}T \ket{\psi} = \braket{\psi'}{\psi'} \ge 0, \ \ \forall \ket{\psi}.
\end{equation}
\noindent Moreover, the trace condition can be achieved by normalization:
\begin{equation}
\frac{T^{\dagger}T}{\trace{T^{\dagger}T}}.
\label{Norm}
\end{equation}
\noindent This type of matrix satisfies the mathematical properties needed for a density matrix.\\
\\
For example, if the state space has dimension $d$, a natural choice for the matrix $T$ in \refe{Norm} is an upper  triangular matrix, given by
\begin{equation}
\label{mt}
T(t) = \left( 
\begin{array}{ccccc}
t_1 & t_{d+1} + i t_{d+2}  & \dots  & \dots & t_{d^2 - 1} + i t_{d^2} \\
 \\
0   & t_2 & t_{d+3} + i t_{d+4} & \ddots & \vdots\\
 \\
0   & 0   & \ddots & \ddots & \vdots \\
 \\
\vdots   &  \ddots  & \ddots & \ddots & t_{3d - 1} + i t_{3d - 2} \\
 \\
0       & \dots & \dots & 0 & t_d
\end{array}
\right).
\end{equation}
\noindent This choice, presented in \cite{james2001}, is motivated by the fact that in a system of $n$ qubits, we need to determine $4^n$ parameters (the Stokes parameters) to estimate the density matrix. Moreover, this choice resembles the Cholesky factor \cite{meyer2001} if only positive diagonal entries are considered. \\
Then, we have a parameterized version of the matrix $\rho$:
\begin{equation}
\label{rhot}
\rho(t) = \frac{T(t)^{\dagger}T(t)}{\trace{T(t)^{\dagger}T(t)}}.
\end{equation}
\noindent \ \\
Using this parameterization the constraints over the matrix are satisfied. Notice this is a continuous and surjective mapping from parameters space $\R_*^{d^2} = \R^{d^2} \setminus \set{0}$ to the density matrices space $\cal{S}$. However, $\rho(t)$ is not one-to-one. For example, consider the matrix:
$$
\rho = \left( 
\begin{array}{cc}
\label{matriz}
1/2 & 0 \\
0 & 1/2 
\end{array}
\right).
$$
\noindent By inspection of \refe{rhot}, it is easy to find at least four solutions $t = (\ \pm 1/\sqrt{2}, \pm 1/\sqrt{2}, 0, 0 )^{\dagger}$. Moreover, note that any nonzero multiple of $t$, $\alpha t \ \ \forall \alpha \ne 0$, leads to the same matrix.\\
\\
The parameterization \refe{rhot} ensures positivity and unit trace, but now we need to minimize the function $F(\rho(t))$ with respect to the new parameters $t_i$'s. Of course, the structure of this function depends on the probability distribution assumption for the experimental uncertainties.\\
\\
Commonly, $F(\rho)$ is a convex function of $\rho$, for $\rho \in {\cal S}$. However, after the parameterization \refe{rhot}, the new objective function $F(\rho(t))$ becomes a nonconvex function of $t$ with several local minima. \\
\\
To illustrate this fact consider the photonic state tomography as treated in \cite{altepeter2005}. In this case the noise is assumed to be Gaussian, as well as in many other physical systems. Then, the probability of getting the observed data $n$, given $\rho(t)$, is:
\begin{equation}
P(n \,|\,\rho(t)) = \frac{1}{N_0}\prod_{\mu} \exp\left[- \frac{(n_{\mu} - \bar{n}_{\mu})^2}{2 \bar{\sigma}_{\mu}^2}  \right],
\label{Gauss}
\end{equation}
\noindent where $\bar{n}_{\mu}$ is the expected value, $\bar{\sigma}_{\mu}$ is the standard deviation (approximately $\sqrt{\bar{n}_{\mu}}$) for  the number of times the $\mu$-th measurement result occurs, and $N_0$ is a normalization factor.\\
According to \refe{expectation},
\begin{equation}
\bar{n}_{\mu} = N\trace{O_{\mu} \rho(t)}.
\label{average}
\end{equation}
\noindent So, we have
\begin{equation}
P(n \, | \rho(t)) = \frac{1}{N_0}\prod_{\mu} \exp\left[- \frac{(N\trace{O_{\mu} \rho(t)} - n_{\mu})^2}{2 \, N\trace{O_{\mu} \rho(t)}}  \right],
\label{P}
\end{equation}
\noindent and to maximize this function is sufficient to minimize the negative log-likelihood:
\begin{equation}
\label{likefun}
F(\rho(t)) = \sum_{\mu}  \frac{[ \trace{O_{\mu} \rho(t)} - f_{\mu} ]^2}{2 \, \trace{O_{\mu} \rho(t)}} = \frac{1}{2} \sum_{\mu} \left( \frac{\trace{O_{\mu} \rho(t)} - f_{\mu}}{\sqrt{\trace{O_{\mu} \rho(t)}}} \right)^2.
\end{equation}
\noindent This is a nonconvex function of $t$, with several local minimizers. For instance, if we consider the one qubit tomography case for the polarization of a single photon, the matrix $\rho(t)$ is:
\begin{equation}
\label{mt1}
\rho(t) =
\frac{1}{t_1^2 + t_2^2 + t_3^2 + t_4^2}
\left(
\begin{array}{cc}
t_1^2 & t_1 t_3 + i\, t_1 t_4 \\
t_1 t_3 - i\, t_1 t_4 & t_2^2  + t_3^2 + t_4^2
\end{array}
\right).
\end{equation}
\noindent Making projective measurements based on $\set{ \ket{H},\ket{V},\ket{D},\ket{R}}$  \cite{altepeter2005, james2001}, which correspond respectively to the linear polarization states horizontal, vertical, diagonal and to the right circular polarization state, we have:
\begin{eqnarray}
\trace{\ket{H}\bra{H} \rho} & = & \bra{H} \rho(t) \ket{H}  =  \frac{t_1^2}{t_1^2 + t_2^2 + t_3^2 + t_4^2}, \\ \nonumber
\trace{\ket{V}\bra{V} \rho} & = & \bra{V} \rho(t) \ket{V}  =  \frac{t_2^2 + t_3^2 + t_4^2}{t_1^2 + t_2^2 + t_3^2 + t_4^2}, \\ \nonumber
\trace{\ket{D}\bra{D} \rho} & = & \bra{D} \rho(t) \ket{D}  =  \frac{1}{2}\left(1 + \frac{2t_1 t_3}{t_1^2 + t_2^2 + t_3^2 + t_4^2} \right), \\ \nonumber
\trace{\ket{R}\bra{R} \rho} & = & \bra{R} \rho(t) \ket{R}  =  \frac{1}{2}\left(1 - \frac{2t_1 t_4}{t_1^2 + t_2^2 + t_3^2 + t_4^2} \right).
\end{eqnarray}
Substituting these components in expression \refe{likefun}, clearly we obtain a nonconvex function $F(t)$. The  examples of Section \ref{num_ex} illustrate this fact.\\
\\
Since $F(\rho(t))$ is a nonconvex function of $t$, we could have problems in the optimization process. If local, non-global minimizers exist, the algorithm could find a local minimizer $t^*$ that is not a global one and the corresponding density matrix $\rho(t^*)$ might not be the MLE $\hat{\rho}$.\\
\\
Some papers in the field \cite{james2001, altepeter2005, usami2003} consider the existence of several local minima which are not global.
In this case, a good initial guess must be provided, so that the optimization procedure can achieve the global minimum. In order to overcome this difficult, those authors propose smart starting points for the optimization method, hoping that these points will lead to  convergence for a global minimum. For example, in \cite{james2001}, the initial guess is based on $\rho_{LM}$ obtained from the linear inversion. However, calculation of this starting point, requires
solution of a linear system, eigenvalues estimation, and matrix factorization. This represents a high computational cost and becomes prohibitive for large systems. Moreover, there is no guarantee that the global minimum is achieved.\\
\\
In Section \ref{main} we prove that, contrary to what was believed, all local minimizers $t$ of $F(\rho(t))$ are equivalent, in the sense that they all lead to the same reconstructed density matrix $\rho(t)$. As $F(\rho)$ we consider, not only the function \refe{likefun} of photonic tomography case, but any convex function of $\rho$ in ${\cal S}$. Indeed, we exploit the mathematical properties of the mapping $\rho(t)$, defined by \refe{rhot}, to prove that all local minimizers of $F(\rho(t))$ are equivalent. \\
\\

First, the Section \ref{theory} addresses the mathematical theory needed to prove that all local minimizers of $F(\rho(t))$, for $t \in \R_{**}^{d^2}$, are in fact global.

\section{The mapping $\rho(t)$ and the feasible set ${\cal S}$}\label{theory}

We start this section by considering the general problem of
optimizing the convex function $f(x)$, such that $x \in \Omega$, with $\Omega$
also convex, given the parameterization $x(t)$. We develop
sufficient conditions to prove the equivalence between local minimizers.
Then we address the optimization of $F(\rho(t))$ as a particular
case. 
\\
\\
Consider the convex optimization problem
\begin{equation}
\begin{aligned}
\min_{x} \ \ \ & f(x) \\
\text{s.a} \ \ \ & x \in \Omega,
\end{aligned}
\label{convexprob}
\end{equation}
where $f$ is convex on the convex set $\Omega \subset  \Rn$.\\
\\
Let $x=x(t)$ be a continuous and surjective mapping $x: D \rightarrow \Omega$, $D \subset \Rn$ and consider the associated unconstrained optimization problem:
\begin{equation}
\begin{aligned}
\min_{t} \ \ \ & \phi(t) = f(x(t)). \\
\end{aligned}
\label{tprob}
\end{equation}
\noindent It is clear that if $t^*$ is a global minimizer of \refe{tprob}, then $x^* = x(t^*)$ is also a global minimizer of \refe{convexprob}. In fact,  $\phi(t^*) \le \phi(t) \ \forall t$, that is, $f(x(t^*)) \le f(x(t)) \ \forall t$, and since $x(t)$ is surjective, this implies that $f(x^*) \le f(x), \ \forall x \in \Omega$.\\
\ \\
However, in general, optimization methods just can find local minimizers, and some local minimizers of  \refe{tprob} could not be global. 
Therefore, we need that a stronger assumption holds for the mapping $x(t)$ in order to show that all local minimizers of \refe{tprob} are global. 
 
If $x(t)$ is an homeomorphism, onto $\Omega$, in the subsets $D_{\alpha} \subset D$ such that $D = \cup_{\alpha} D_{\alpha}$ then we can prove the desired result. This condition is weaker than requiring that $x(t)$ be a homeomorphism, and it is easier to hold for some mappings, in particular for mapping $\rho(t)$ in QST. \\
\\
\noindent {\bf Assumption A1.} For the mapping $x(t)$ $x: D \rightarrow \Omega$, there are sets $D_{\alpha} \subset D$ such that $D = \cup_{\alpha} D_{\alpha}$ and, for each $D_{\alpha}$, $x: D_{\alpha} \rightarrow \Omega$ is an homeomorphism. \\

\begin{teo} \label{mainteo} Let $x(t)$, $x: D \rightarrow \Omega$ be continuous and surjective. If Assumption A1 holds, then any local minimizer of \refe{tprob} is global.
\end{teo}
\begin{proof} Let $t^*$ be a global minimizer of \refe{tprob} and suppose there is a local, non-global, minimizer $\hat{t}$. Then, $f(x(\hat{t})) \le f(x(t)) \ \forall t \in B(\hat{t},\delta) \inter D$, but  $f(x(t^*))<f(x(\hat{t}))$. \\ Let $x^*=x(t^*)$ and $\hat{x} = x(\hat{t})$. Using the convexity of $f$ in $\Omega$, we have:
\begin{equation*}
f(\lambda x^* + (1-\lambda)\hat{x}) \le \lambda f(x^*) + (1-\lambda)f(\hat{x}) < \lambda f(\hat{x}) + (1-\lambda)f(\hat{x}) = f(\hat{x}), \ \ \ \forall \lambda \in (0,1).
\end{equation*}
\noindent Notice that $\tilde{x} = \lambda x^* + (1-\lambda)\hat{x} \in \Omega$, $\forall \lambda \in (0,1)$, since $\Omega$ is a convex set, and for $\lambda$ sufficiently close to zero, $\tilde{x}$ is as close to  $\hat{x}$ as necessary, that is, given $\eps>0$ there exists $\lambda>0$ such that $\tilde{x} \in B(\hat{x},\eps) \inter \Omega$. \\
\\
Since $x(t)$ satisfies Assumption A1,  $x(t)$ is an homeomorphism from $D_{\alpha}$ onto $\Omega$, where $D_{\alpha}$ is such that $\hat{t} \in D_{\alpha}$. Thus, by continuity of the inverse of $x(t)$, given $\delta_0>0$, there exists $\eps(\delta_0)>0$ such that $\tilde{x} \in B(\hat{x},\eps(\delta_0)) \inter \Omega$ implies  $\tilde{t} \in B(\hat{t},\delta_0) \inter D$, where $\tilde{t} \in D_{\alpha}$, $\tilde{t}=x^{-1}(\tilde{x})$. In particular for $\delta_0<\delta$. Therefore, there is $\tilde{t} \in B(\hat{t},\delta) \inter D$  such that $f(x(\tilde{t}))<f(x(\hat{t}))$, which contradicts the fact that $\hat{t}$ is a local minimizer.
\end{proof} 
\ \\
\noindent Now, as a particular case of the above results we will prove the equivalence between local minimizers of $F(\rho(t))$. To do this, we need to show that Assumption A1 holds for $\rho(t)$ defined by \refe{rhot}. Actually, we show that $\rho: \R_{**}^{d^2} \rightarrow \interior{{\cal S}}$ satisfies the Assumption~A1, where $\R_{**}^{d^2} = \set{t \in \R_*^{d^2} \ |\ t_i \ne 0,\ \forall i=1,\dots,d}$ and $\interior{{\cal S}}$ is the relative interior of ${\cal S}$. 
\begin{prop}
\label{propkey}
The Assumption A1 holds for the mapping $\rho: \R_{**}^{d^2} \rightarrow \interior{{\cal S}}$.
\end{prop}
\begin{proof} 
First, we need to define the sets $D_{\alpha,j}$ such that $\R_{**}^{d^2} = \cup_{\alpha,j} D_{\alpha,j}$. The index $j$ is one-to-one with the possible sign permutations of the diagonal elements of $T(t)$ ($2^{d}$ possible sign permutations), that is,
$$
D_{\alpha, j} = \set{t \in \R_{**}^{d^2} \ | \ \norm{t}^2_2 = \alpha \mbox{ and  $t_1,\dots,t_d$ satisfing the $j$-sign configuration}},
$$
which implies that, for $t \in D_{\alpha,j}$, 
$$
\rho(t) = \frac{1}{\alpha}T(t)^{\dagger}T(t).
$$
Cholesky's factorization \cite{meyer2001,golub1996} states that an Hermitian matrix $A$ is positive definite, if and only if there is a unique upper triangular matrix $T$, with positive diagonal elements such that $A = T^{\dagger}T$. We stress that the existence and uniqueness of the upper triangular factor $T$ holds not only for positive diagonal elements, but also for any fixed order of signs of the diagonal elements $T$. Moreover, the triangular factor is continuous because each of its entries is a composition of continuous functions of the entries of $A$, \cite{stweart1997}. \\
\\
Thus, $\rho: D_{\alpha,j} \rightarrow \interior{{\cal S}}$ defines an homeomorphism because it is continuous, bijective, and has continuous inverse, which is just the triangular factor with a fixed choice of signs. Since $\R_{**}^{d^2} = \cup_{\alpha,j} D_{\alpha,j}$, then $\rho:\R_{**}^{d^2} \rightarrow \interior{{\cal S}}$ satisfies Assumption A1.
\end{proof}

\begin{figure}[htbp] 
  \centering
  \includegraphics[scale=0.3]{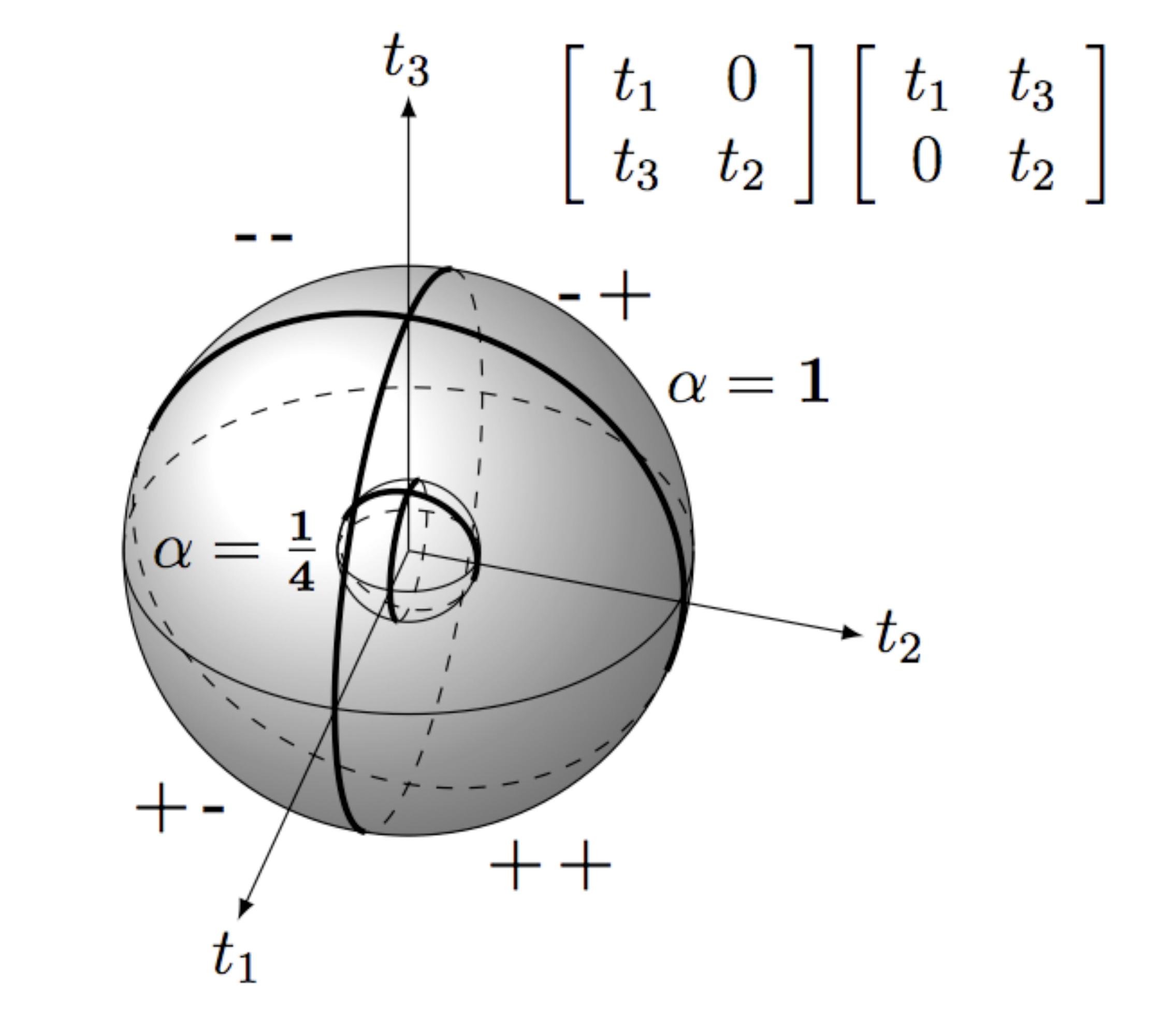}
  \caption{Covering $\R_{**}^{d^2}$ with the $D_{\alpha,j}$ sets.}
  \label{afigura}
\end{figure}

\noindent  Figure \ref{afigura} shows how the sets $D_{\alpha,j}$ are defined in the case of a $2 \times 2$ symmetric matrix.

\section{Main Result}\label{main}
In this section, using the Theorem \ref{mainteo} and Proposition \ref{propkey}, we prove that all local minimizers of the negative log-likelihood $F(\rho(t))$ are in fact global. 
\noindent For any convex function $F(\rho)$, the problem of minimizing $F(\rho)$ on $\cal{S}$ is a convex optimization problem, where local minimizers $\rho^*$ are also global. Indeed, if we have an informationally complete set of measurements, then the minimizer of this convex optimization problem is unique. 

Since we have a convex minimization problem and the mapping $\rho(t)$ satisfies the Assumption A1, as a corollary of Theorem \ref{mainteo} and Proposition \ref{propkey}, we establish the main result of this paper. \\
\begin{cor} \label{local_global}
If $F(\rho)$ is a convex function for $\rho \in {\cal S}$ then all local minimizers of $F(\rho(t))$, for $t \in \R_{**}^{d^2}$, are also global.
\end{cor}
\begin{proof} Since $F(\rho)$ is convex in ${\cal S}$ then it is also convex in $\interior{{\cal S}}$. Thus, minimizing $F(\rho)$ in $\interior{{\cal S}}$ is a convex optimization problem. By Proposition \ref{propkey} the mapping $\rho(t)$, $\rho: \R_{**}^{d^2} \rightarrow \interior{{\cal S}}$, satisfies Assumption A1, then Theorem \ref{mainteo} applies, which implies that all local minimizers of $F(\rho(t))$ in $\R_{**}^{d^2}$ are global.
\end{proof}
\ \\
\noindent The practical result provided by Corollary \ref{local_global}  is that it does not matter what local minimizer  $\hat{t} \in \R_{**}^{d^2}$ is found. All of them are equivalent, in the sense that they lead to the same density matrix $\hat{\rho} = \rho(\hat{t})$ corresponding to the maximum likelihood estimation. \\
\\
Albeit the above result considers $\rho(t)$ restricted to $\R_{**}^{d^2}$ onto $\interior{{\cal S}}$, it is straightforward to show that these are dense sets in $\R_*^{d^2}$ and ${\cal S}$, respectively. Thus, by continuity of $F(\rho)$ and $F(\rho(t))$, we have
$$
\min_{\rho \, \in \, {\cal S}} F(\rho) \  = \inf_{\rho \, \in \, \interior{{\cal S}} } F(\rho) \ \ \ \ \ \mbox{ and } \ \ \  \ \min_{t \, \in \, \R_*^{d^2}} F(\rho(t))\  = \inf_{t \, \in \, \R_{**}^{d^2}} F(\rho(t)).
$$
\noindent Therefore, in practice, it is enough to consider $\R_{**}^{d^2}$ instead of $\R_*^{d^2}$. Indeed, it is enough to consider just one set $D_{\alpha,j}$, for instance, restricting $\norm{t}_2^2 = 1$ and $t_i>0$ for $i=1,\dots,d$. But these constraints become the optimization problem harder than the unconstrained one, and the Corollary \ref{local_global} gives us the guarantee that any local minimizer of the unconstrained problem in $\R_{**}^{d^2}$ can be used.\\
\\
In fact, points belonging to $\R_*^{d^2} \setminus \R_{**}^{d^2}$ correspond to rank deficient density matrices that live in the boundary of ${\cal S}$. From the numerical optimization point of view, solutions in the boundary, in general, are approached only in the limit. That is, numerically, a point in the interior that is  sufficiently close to the true boundary solution is declared as the solution within certain tolerance criteria. Besides, the negative log-likelihood function, for example \refe{likefun}, could not be defined (goes to infinity) in some states in the boundary, depending on the chosen POVM set. From the experimental point of view, matrices in the boundary of ${\cal S}$ correspond to states having zero eigenvalues, for example pure states, that are rarely seen in practice due to either difficulty of realization or by imperfections in the apparatus leading to noisy experimental data.\\
\\
Finally, we stress that the above result remains valid for any convex function $F(\rho)$ over ${\cal S}$. In particular we can apply the result to the negative log-likelihood function $F(\rho)=-\log P(n | \rho)$ because  whether $P(n | \rho)$ is Gaussian or multinomial, the function $F(\rho)$ will be convex. 

\section{Numerical Experience}\label{num_ex}

\begin{figure} 
\vspace*{-3cm}
  \includegraphics[bb=218 535 396 708,width=2cm,height=2cm,keepaspectratio]{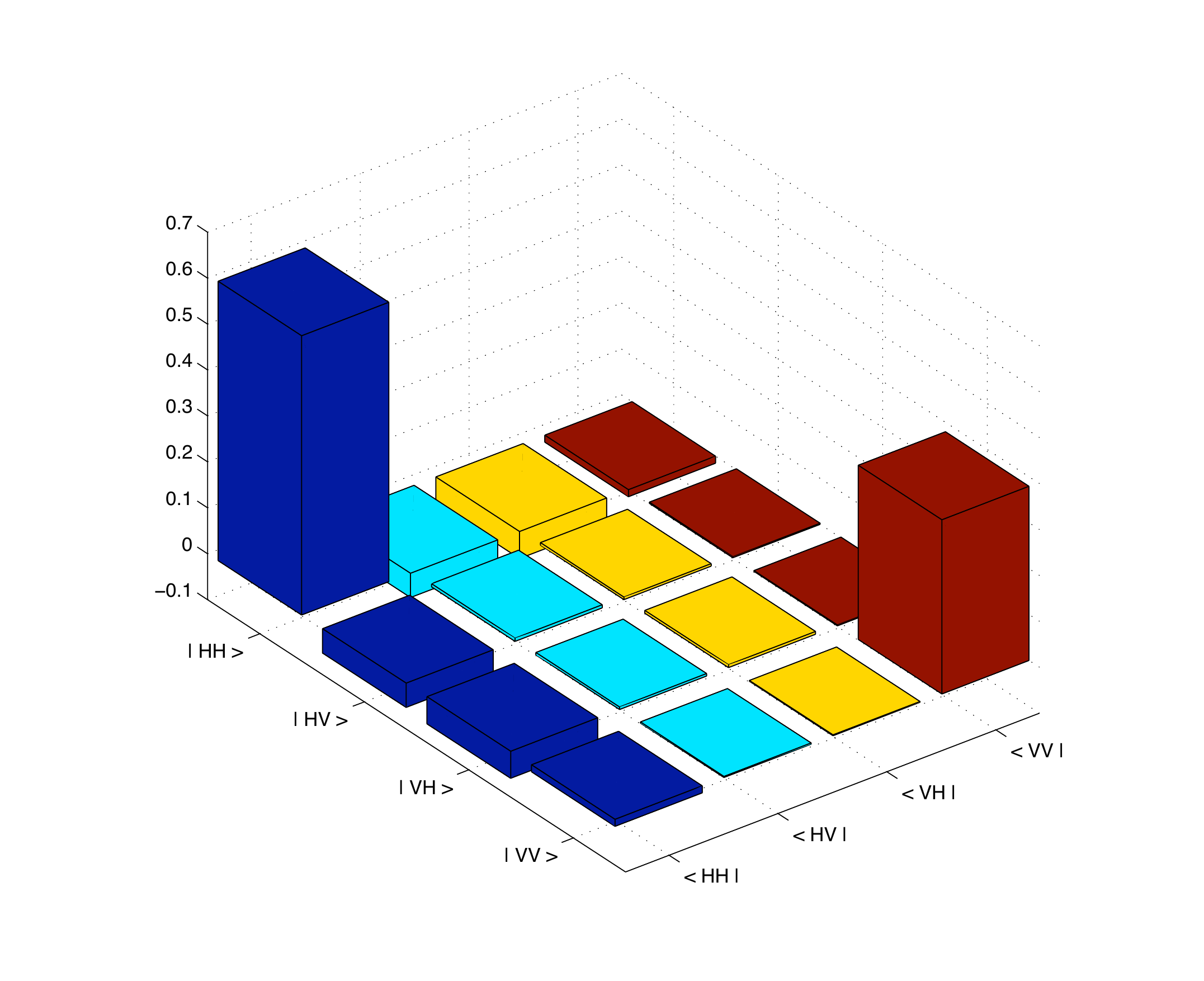} \hspace{4cm}
\hspace*{1cm}
  \includegraphics[bb=218 535 396 708,width=2cm,height=2cm,keepaspectratio]{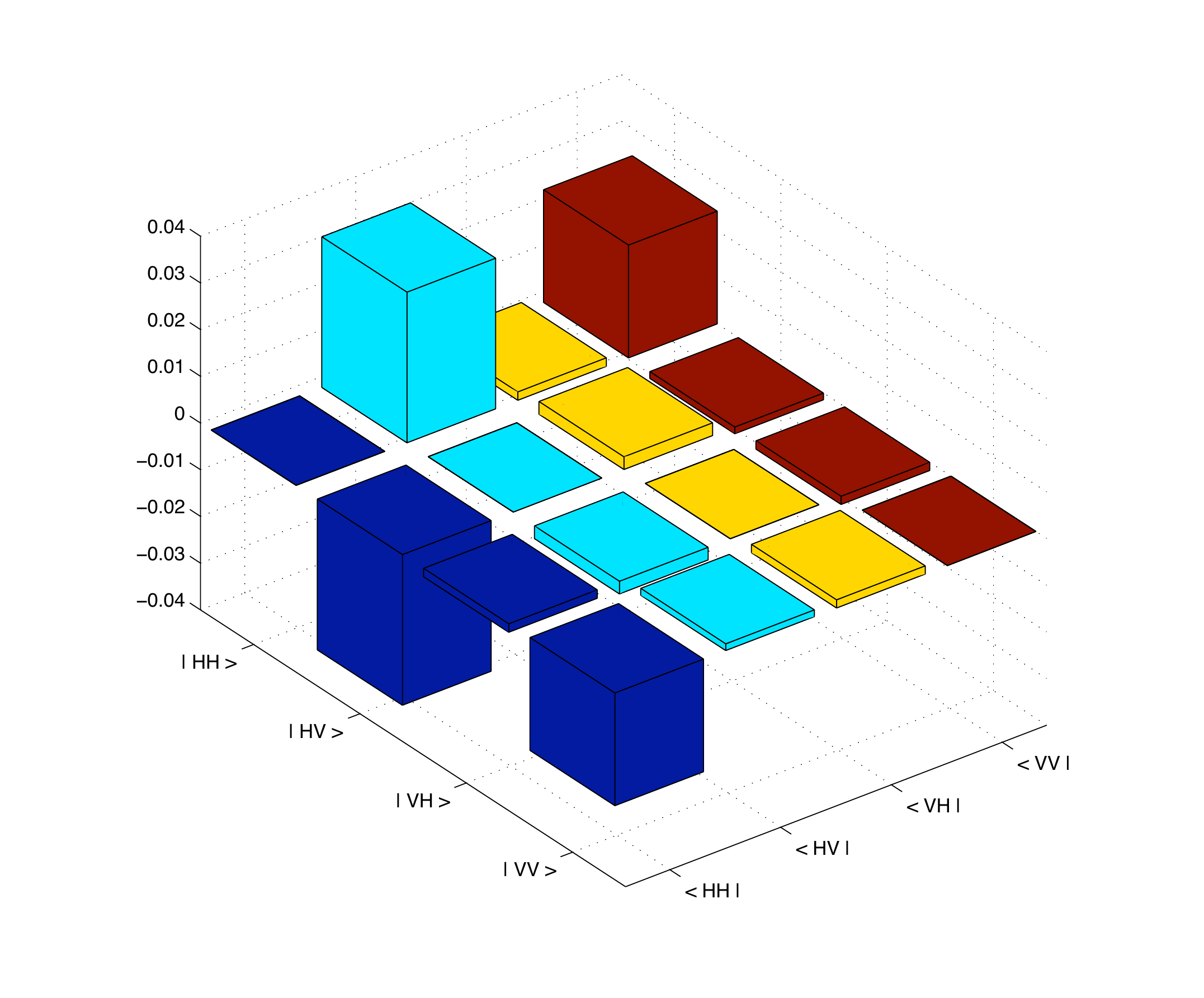} \vspace{6cm}
  \caption{Real (left) and imaginary (right) parts of density matrix of Example 2 using the derivative-free solver {\tt fminsearch} of {\tt MATLAB}.}
  \label{dmfms}
\end{figure}

\noindent We pointed out that in some papers \cite{altepeter2005, james2001, usami2003}, the possibility of finding local minimizers that are not global, in the minimization of function \refe{likefun}, is considered. However in Section \ref{main} we showed that all local minimizers of $F(\rho(t))$ in $\R_{**}^{d^2}$ are also global. What may have occurred in those cases was the stagnation of certain methods which did not have global convergence property \cite{luenberger2003, nocedal1999} (this property should not be confused with convergence to global minimizers). 
\begin{defi}
We say that an algorithm for the optimization problem
$$
\min_{x \in \Rn} F(x)
$$
\noindent has global convergence property to stationary points (\ $\nabla F(x) = 0$\ ), if each accumulation point of the sequence of iterates $\set{x_k}_{k=1}^{\infty}$ generated by this algorithm is a stationary point of $F(x)$. 
\end{defi}
\noindent Some unconstrained optimization methods, for instance the known Nelder-Mead method \cite{csv2009,nm1965}, have no global convergence property, that is, a stagnation point could be not even
stationary. Also, commercial packages implement some methods that have no global convergence: for example, {\tt fminsearch} of {\tt MATLAB}, which implements the Nelder-Mead method, and {\tt FindMinimum} of {\tt Mathematica}, which implements the Powell's method \cite{powell1964}. Both of these algorithms, used in quantum state tomography,  may be stopped prematurely, before reaching a local minimizer or a stationary point.\\
\\
Even methods that have global convergence property can still stop prematurely due to numerical difficulties. 
Thus, it is important to investigate which stopping criterion was triggered, in order to avoid wrong conclusions. In practice, we usually hope that an optimization algorithm could find, at least, an  $\eps-$stationary point $x^ *$, that is, $\norm{\nabla F(x^*)}<\eps$. However, the optimization problem is sometimes not well scaled, or so difficult, that another stopping criteria must be employed to avoid that the iterative algorithm runs indefinitely. Normally, stagnation criteria like small changes in variables $\norm{x_{k+1} - x_k} < \eps_x$ or in objective function $\abs{F(x_{k+1}) - F(x_k)}< \eps_F$ are used and upper bounds are fixed for the maximum number of iterations or function evaluations. Since we prefer the $\eps-$stationarity criterion, the other tolerances should be smaller than $\eps$, for example $\eps_x = \eps_F = \eps^2$ when $0< \eps \ll 1$. \\
\\
Another undesirable consequence of parameterization \refe{rhot} is that when $\norm{t} \rightarrow \infty$,  $\norm{\nabla_t F(\rho(t))} \rightarrow 0$. So, optimization algorithms using as stopping criterion $\norm{\nabla F(\rho(t))} < \eps$, can stop in $t$ points having large $\norm{t}$, even if $t$ is not a minimizer of \refe{likefun}. Thus, artificial bounds on the variables may be used in order to avoid this situation.  \\
\\
The next examples show what can go wrong in QST using inappropriate methods, without global convergence or not numerically stable. For these examples we fix the tolerances $\eps_x=\eps_F=10^{-8}$ and the maximum number of iterations and function evaluations $M=2 \times 200 \times n$, where $n$ is the number of variables of the optimization problem and $200 \times n$ is the default limit used in ${\tt fminsearch}$ routine of ${\tt MATLAB}$. One can set these tolerances using the {\tt optimset} routine. Unfortunately, {\tt MATLAB} also uses the relative function change tolerance $\eps_F$ as the gradient norm tolerance $\eps$, so in the examples 1 and 2, $\eps=10^{-8}$.  \\
\\
{\bf Example 1.} Consider $\rho = \ket{H}\bra{H}$ as the true state that we want to identify with tomography. The expected values for projective measurements based on $\set{\ket{H},\ket{V},\ket{D},\ket{R}}$, perturbed by some kind of error (computationally generated based on Gaussian distribution), plays the role of normalized frequencies in an experiment: $f_H = 0.9990$, $f_V = 0.0002$, $f_D = 0.4995$, $f_R = 0.4994$. \noindent Using {\tt fminsearch} of {\tt MATLAB R2009a}, with starting point $t_0 = (-0.0001,\, 0.999,\, 0.001,\, 0.999)^{T}$, we recover the following matrix:
$$
\rho = \left[
\begin{array}{ll}
   0.2219        &    -0.3634 + 0.0002i \\
  -0.3634 - 0.0002i &  0.7781 
\end{array} \right].
$$
\noindent The optimization procedure stopped after $k=595$ iterations, with objective function value $\bar{F}=2.2316$, and gradient norm $\norm{\nabla \bar{F}}=0.0013$. The stopping criterion verified was $\norm{\bar{x} - x_l}_{\infty} \le \eps_x$ and $\abs{\bar{f} - f_l} \le \eps_F$, where $\bar{f} = f(\bar{x})$ is the function value in the best point $\bar{x}$ and $f_l$'s are the function values on the other simplex points $x_l$'s.  \\
On the other hand, we used {\tt lsqnonlin} routine for nonlinear least squares minimization, also of {\tt MATLAB R2009a}, regarding the special structure of function \refe{likefun}. Providing the partial derivatives to compose the Jacobian matrix, the command runs a trust-region-reflective algorithm \cite{coleman1996}. Using the same starting point, we get:
$$
\rho = \left[
\begin{array}{ll}
0.9998       &     -0.0005 + 0.0006i \\
  -0.0005 - 0.0006i  &  0.0002 
\end{array} \right],
$$ 
\noindent with optimization outputs $\bar{F} = 3.2 \times 10^{-7}, \norm{ \nabla \bar{F}} = 5.3024 \times 10^{-10}$, after $k=22$ iterations. Notice the difference between the gradient norms in these two cases. Clearly, the first does not correspond to a stationary point, nor a local minimizer. \\
\\
\noindent Another important fact is the non-unique parameterization of $\rho(t)$. Consider the same data of the above example. Using {\tt lsqnonlin} from the starting point $t_0 =(1,\, 0.001,\, 0,\, 0)^T$, we obtain the solution
$$
\hat{t} = (0.999,\, 0.0141,\, -0.0005,\, 0.0006)^T,
$$
which gives the matrix
$$
\rho(\hat{t}) = \left[
\begin{array}{ll}
0.9998       &     -0.0005 + 0.0006i \\
  -0.0005 - 0.0006i &  0.0002   
\end{array} \right].
$$
Starting from another initial point $t'_0 =(-1,\, -0.001,\, 0,\,  0)^T$, the solution is
$$
t' = (-0.999,\, -0.0141,\, 0.0005,\, -0.0006)^T,
$$
which gives the same matrix
$$
\rho(t') = \left[
\begin{array}{ll}
0.9998       &     -0.0005 + 0.0006i \\
  -0.0005 - 0.0006i &  0.0002  
\end{array} \right].
$$
\noindent Both $\hat{t}$ and $t'$  satisfy the numerical stationarity criterion $\norm{\nabla F} <  \eps$. Despite the distinct local minimizers, the recovered matrix is the same.\\
\\
\noindent {\bf Example 2.} Consider the following experimental data  extracted from a complete set of measurements in the polarization degrees of freedom of two photons generated in a Parametric Down Conversion Process, where the prepared state is nearly pure:
$$
\begin{array}{lcrlcrlcrlcr}
n_1 & = & 3043 & n_5 & = &        1546 & n_9 & = &        1556 & n_{13} & = &        1070 \\
n_2 & = &          32 & n_6 & = &        1283  & n_{10} & = &         122 & n_{14} & = &        1048 \\ 
n_3 & = &        2159 & n_7 & = &         938 & n_{11} & = &        1271 & n_{15} & = &        1611 \\
n_4 & = &          19 & n_8 & = &        1595 & n_{12} & = &        1621 & n_{16} & = &         114. \\
\end{array}
$$
First, we use {\tt fminsearch} of {\tt MATLAB R2009a}, with initial guess $t_i = 1/\sqrt{d}$ for $i \le d$ and  $t_i=0$, for $i>d$ (corresponding to the maximally mixed state $,I/d$). The algorithm stops because the maximum number of function evaluation ($2 \times 16 \times 200 = 6400$) is reached. The recovered density matrix is shown in Figure \ref{dmfms}, and presents purity  $\trace{\rho_1^2} = 0.5298$.

\begin{figure} 
\vspace{-3cm}
  \includegraphics[bb=218 535 396 708,width=2cm,height=2cm,keepaspectratio]{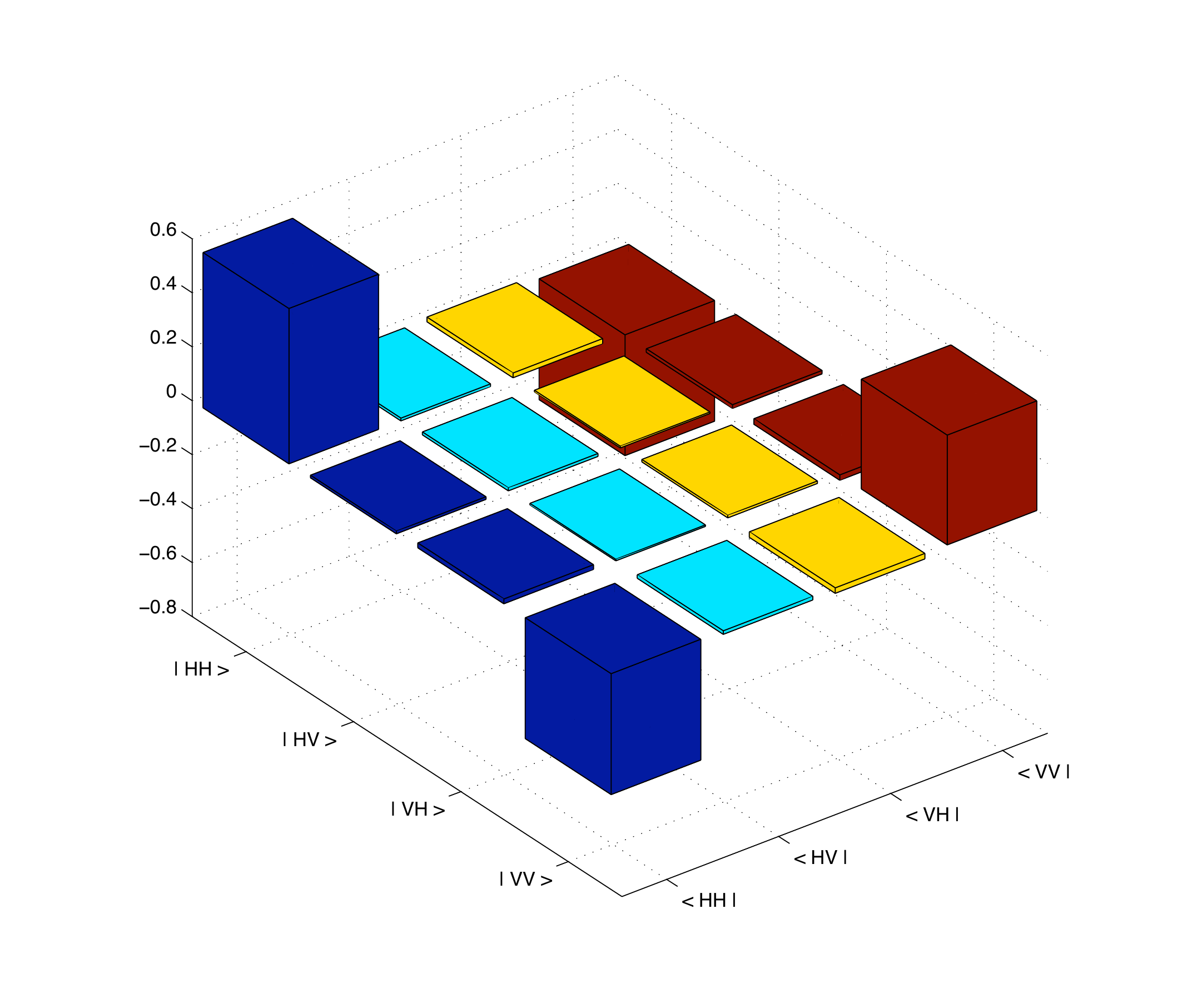} \hspace{4cm}
\hspace*{1cm}
  \includegraphics[bb=218 535 396 708,width=2cm,height=2cm,keepaspectratio]{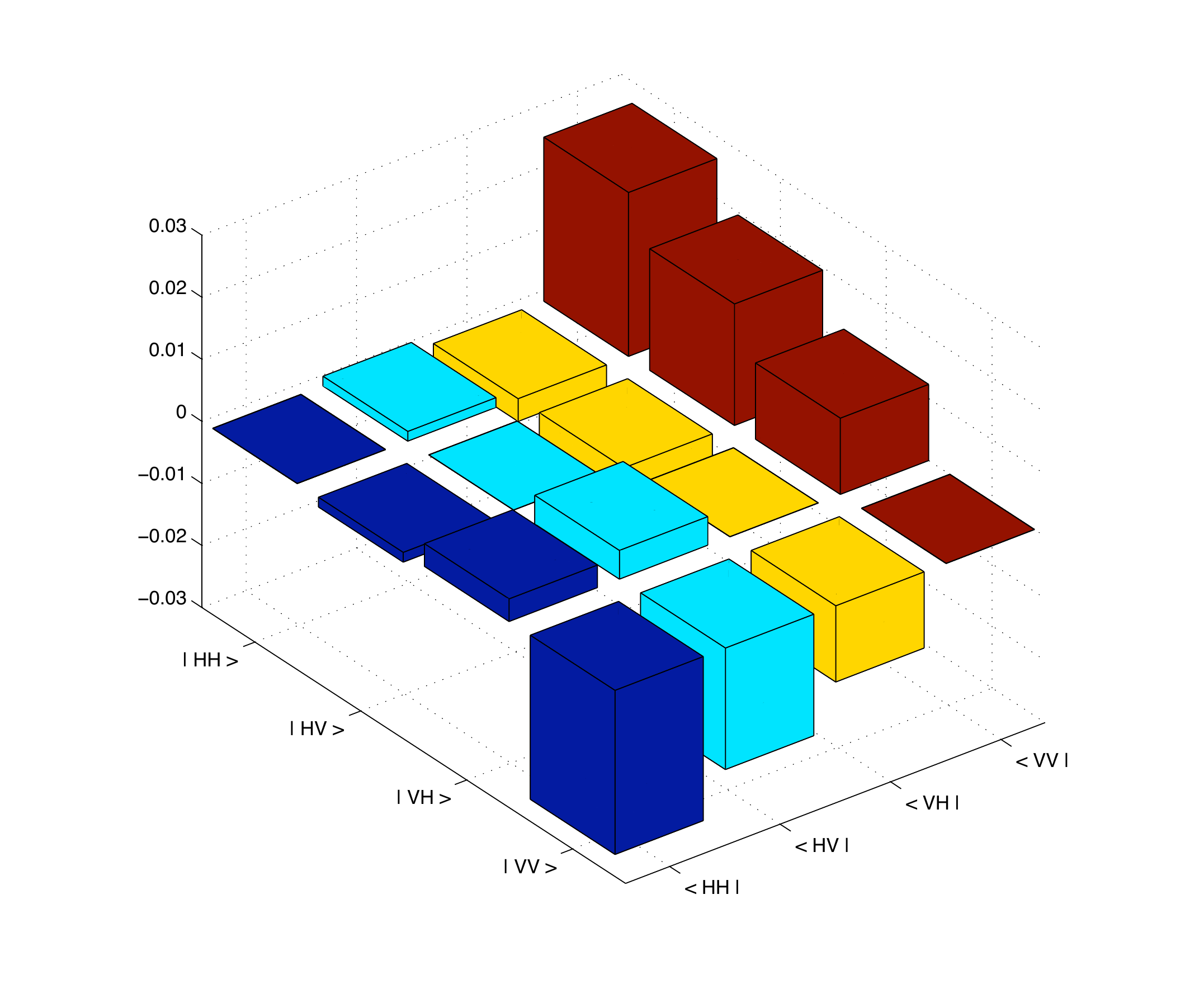} \vspace{6cm}
  \caption{Real (left) and imaginary (right) parts of density matrix of Example 2 using the nonlinear least-squares solver {\tt lsqnonlin} of {\tt MATLAB}.}
  \label{dmsnls}
\end{figure}

\noindent 
Second, using {\tt lsqnonlin} of {\tt MATLAB R2009a}, with the same initial point, we get the density matrix of Figure \ref{dmsnls}, with purity $\trace{\rho_2^2} = 0.9008$.

\vspace{.5cm}
\noindent {\bf Example 3.} To conclude the numerical examples, and to remark the meaning of Corollary  \ref{local_global}, consider the state $\rho = \ket{R}\bra{R}$, and its normalized frequencies
$f_H  = 0.5$, $f_V = 0.5$, $f_D = 0.5$ and $f_R = 1$. The theory presented in Section \ref{main} tell us that the solution of minimizing $F(\rho(t))$ in $D_{\alpha,j}$ is unique, for each $\alpha$ and $j$. So, if we include the constraint $\norm{t}_2^2=1$, that is, fixing $\alpha=1$ and choosing one fixed sign ordering among $++$,$+-$,$-+$,$--$, in the case of $2 \times 2$ matrices, then we have exactly one solution. To test this numerically, we define four optimization problems of minimizing $F(\rho(t))$ subject to $\norm{t}_2^2=1$ and for each problem the bound constraints: $t_1>0$ and $t_2>0$, $t_1>0$ and $t_2<0$, $t_1<0$ and $t_2>0$, $t_1<0$ and $t_2<0$. For each problem, we use a multistart procedure to generate $23$ random initial points with components in the interval $[-1,1]$, which were used by {\tt fmincon} routine to solve each constrained minimization problem. Inside each multistart procedure, we consider the tolerances $\eps_x=\eps_F=10^{-12}$ and that a problem is solved only if the first order optimality measure is less than $\eps=10^{-6}$, discarding solutions that do not match this criterion. We also considered, for each problem, a list of solutions, and after a new candidate solution $\hat{t}$ was obtained we include it in the list only if $\norm{\hat{t} - t}_2>10^{-2}$. For each choice of signs, each list had only one member as shown in Table \ref{tab1}. 
\begin{table}[!h]
\begin{tabular}{c|cccc}
\  & $++$ & $+-$ & $-+$ & $--$ \\ \hline
$t_1$ & 0.7071 & 0.7071 & -0.7071 & -0.7071 \\
$t_2$ & 0.0024 & -0.0000 & 0.0001 & -0.0002\\
$t_3$ & 0.0000 & 0.0000 & -0.0000 & 0.0000\\
$t_4$ & -0.7071 & -0.7071 & 0.7071 & 0.7071
\end{tabular}
\caption{Solutions for each ordering of signs, after 23 random initial points.}
\label{tab1}
\end{table}
This numerical experiment confirms what was expected by the theory developed in Sections \ref{theory} and \ref{main}. \\
\\
\noindent These examples clarify the nature of local minimizers of $F(\rho(t))$ and illustrate some mistakes that can be made in the minimization of negative log-likelihood function when calculating the maximum likelihood estimation, stressing the importance of good theoretical and numerically stable optimization methods, as well as the analysis of stopping criteria and tolerances.

\section{Conclusion}\label{conclusion}

We show that the MLE using the parameterization approach \refe{rhot}, when used with appropriate numerical optimization methods, is a reliable procedure to estimate the density matrix in Quantum State Tomography (QST). We prove that even though this approach implies an unconstrained minimization problem with several possible minimizers, all of them are global. Thus, if one uses a globally convergent and numerically stable minimization method, a local solution, that is also global, will be found regardless of any initial guess. This has important practical consequences, because as discussed, the utilization of smart initial guesses represents an additional computational cost, and therefore becomes prohibitive for quantum state reconstruction of a large number of qubits. \\
\\
Moreover, contrary to what was assumed by some previous papers, our examples show that failures in the estimation of the optimal density matrix can be mostly due to the use of inappropriate minimization methods that do not have global convergence property or suffer from numerical instability issues. This fact has probably been  confused with the existence of local and non-global minimizers associated with the minimization process of MLE. Therefore, care must be taken of the stopping criteria and tolerances. \\
\\
Finally, our results regarding the parameterization $\rho(t)$ and local minimizers of $F(\rho(t))$ remain valid for any convex function $F(\rho)$. We believe that the discussion presented in this work is of interest for most of the quantum optics and quantum information community, and helps to improve the practical procedure of QST. 

\section*{Acknowledgements}
This work was partially supported by the Brazilian research agencies  CNPq, CAPES, FAPESP, FAPERJ and the National Institute for Science and Technology of Quantum Information. We acknowledge Scott Glancy for a discussion about the generality of our results. OJF is funded by
Consejo Nacional de Ciencia y Tecnologia Mexico.

\bibliographystyle{ieeetr}
\bibliography{TOMOGRAPHY_QIC_REVIEW}

\begin{thebibliography}{10}

\bibitem{james2001}
D.~James, P.~Kwiat, W.~Munro, and A.~White, ``Measurement of qubits,'' {\em
  Physical Review A}, vol.~64, no.~5, p.~052312, 2001.

\bibitem{nc2000}
M.~Nielsen and I.~Chuang, {\em Quantum Computation and Quantum Information}.
\newblock Cambridge: Cambridge University Press, 1~ed., 2000.

\bibitem{clavor2011}
T.~Evangelista, C.~Lavor, and W.~R.~M. Rabelo, ``A new method to calculate the
  inconclusive coefficients in the quantum state discrimination,'' {\em
  International Journal of Modern Physics C}, vol.~22, no.~02, pp.~95--105,
  2011.

\bibitem{wz1982}
W.~Wootters and W.~Zurek, ``A single quantum cannot be cloned,'' {\em Nature},
  vol.~299, no.~5886, pp.~802--803, 1982.

\bibitem{vr1989}
K.~Vogel and H.~Risken, ``Determination of quasiprobability distributions in
  terms of probability distributions for the rotated quadrature phase,'' {\em
  Phys. Rev. A; Physical Review A}, vol.~40, no.~5, pp.~2847--2849, 1989.

\bibitem{cb2001}
G.~G.~C. Casella and R.~Berger, {\em Statistical Inference}.
\newblock Duxbury Press, 2~ed., 2001.

\bibitem{lc1998}
E.~Lehmann and G.~Casella, {\em Theory of Point Estimation}.
\newblock Springer, 2nd~ed., 1998.

\bibitem{hradil1997}
Z.~Hradil, ``Quantum-state estimation,'' {\em Physical Review A}, vol.~55,
  no.~3, pp.~R1561--R1564, 1997.

\bibitem{kaznady2009}
M.~Kaznady, ``Numerical strategies for quantum tomography: Alternatives to full
  optimization",'' {\em Phys. Rev. A; Physical Review A}, vol.~79, no.~2, 2009.

\bibitem{rehacek2007}
J.~{\v R}eh{\'a}{\v c}ek, Z.~Hradil, E.~Knill, and A.~Lvovsky, ``{Diluted
  maximum-likelihood algorithm for quantum tomography},'' {\em Physical Review
  A}, vol.~75, no.~4, p.~042108, 2007.

\bibitem{blume2010}
R.~Blume-Kohout, ``Optimal, reliable estimation of quantum states,'' {\em New
  Journal of Physics}, vol.~12, no.~4, p.~043034, 2010.

\bibitem{luenberger2003}
D.~Luenberger, {\em Linear and Nonlinear Programming, Second Edition}.
\newblock Springer, 2nd~ed., 2003.

\bibitem{nocedal1999}
J.~Nocedal and S.~Wright, {\em Numerical optimization}.
\newblock Springer, 1999.

\bibitem{altepeter2005}
J.~Altepeter, E.~Jeffrey, and P.~Kwiat, ``Photonic state tomography,'' vol.~52
  of {\em Advances In Atomic, Molecular, and Optical Physics}, pp.~105 -- 159,
  Academic Press, 2005.

\bibitem{usami2003}
K.~Usami, ``Accuracy of quantum-state estimation utilizing akaike's information
  criterion",'' {\em Phys. Rev. A; Physical Review A}, vol.~68, no.~2, 2003.

\bibitem{boyd2004}
S.~Boyd and L.~Vandenberghe, {\em Convex Optimization}.
\newblock Cambridge University Press, 2004.

\bibitem{todd2001}
M.~Todd, ``Semidefinite optimization,'' {\em Acta Numerica}, vol.~10,
  pp.~515--560, 2001.

\bibitem{meyer2001}
C.~Meyer, {\em Matrix Analysis and Applied Linear Algebra}.
\newblock SIAM: Society for Industrial and Applied Mathematics, 2001.

\bibitem{golub1996}
G.~Golub and C.~Van~Loan, {\em Matrix computations}.
\newblock Johns Hopkins University Press, 3~ed., 1996.

\bibitem{stweart1997}
G.~W. Stewart, ``On the perturbation of {$LU$} and {C}holesky factors,'' {\em
  IMA J. Numer. Anal.}, vol.~17, no.~1, pp.~1--6, 1997.

\bibitem{csv2009}
A.~Conn, K.~Scheinberg, and L.~Vicente, {\em Introduction to Derivative-Free
  Optimization}.
\newblock Society for Industrial and Applied Mathematics, 2009.

\bibitem{nm1965}
J.~Nelder and R.~Mead, ``A simplex method for function minimization,'' {\em The
  Computer Journal}, vol.~7, no.~4, pp.~308--313, 1965.

\bibitem{powell1964}
M.~Powell, ``An efficient method for finding the minimum of a function of
  several variables without calculating derivatives,'' {\em The Computer
  Journal}, vol.~7, no.~2, pp.~155--162, 1964.

\bibitem{coleman1996}
T.~F. Coleman and Y.~Li, ``An interior trust region approach for nonlinear
  minimization subject to bounds,'' {\em SIAM J. on Optimization}, vol.~6,
  no.~2, pp.~418--445, 1996.

\end{thebibliography}

\end{document}